\def\PRL{{ Phys. Rev. Lett.\ }\/}
\def\PRB{{ Phys. Rev. B\ }\/}
\def\PRX{{ Phys. Rev. X\ }\/}

\makeatletter

\newcommand{\Rmnum}[1]{\expandafter\@slowromancap\romannumeral #1@}

\makeatother

\makeatletter
\newcommand*\env@matrix[1][*\c@MaxMatrixCols c]{%
  \hskip -\arraycolsep
  \let\@ifnextchar\new@ifnextchar
  \array{#1}}
\makeatother

\documentclass[aps,prb,showpacs,superscriptaddress,a4paper,twocolumn]{revtex4-2}
\usepackage{amsmath}
\usepackage{float}
\usepackage{color}

\usepackage{mathtools}
\usepackage{textcmds}
\usepackage{braket}

\usepackage[normalem]{ulem}
\usepackage{soul}
\usepackage{amssymb}
\usepackage{tabularx}

\usepackage{amsfonts}
\usepackage{amssymb}
\usepackage{graphicx}
\begin {document}


\title{Quantum Valley and Sub-valley Hall Effect in the Large Angle Twisted Bilayer Graphene} 

\author{Chiranjit Mondal}
\thanks{These authors contributed equally}
\author{Rasoul Ghadimi}
\thanks{These authors contributed equally}

\affiliation{Department of Physics and Astronomy, Seoul National University, Seoul 08826, Korea}
\affiliation{Center for Theoretical Physics (CTP), Seoul National University, Seoul 08826, Korea}

\author{Bohm-Jung Yang}
\email{bjyang@snu.ac.kr}

\affiliation{Department of Physics and Astronomy, Seoul National University, Seoul 08826, Korea}
\affiliation{Center for Theoretical Physics (CTP), Seoul National University, Seoul 08826, Korea}
\affiliation{Center for Correlated Electron Systems, Institute for Basic Science (IBS), Seoul 08826, Korea}

\date{\today}

\begin{abstract}
We study the quantum valley Hall effect and related domain wall modes in twisted bilayer graphene at a large commensurate angle.
Due to the quantum valley and sub-valley Hall effect, a small deviation from the commensurate angle generates two-dimensional conducting network patterns composed of one-dimensional domain walls, which can host non-Fermi liquid behavior within an accessible temperature range.
The domain wall modes can be manipulated using an external electric field and layer shifting, manifesting the physics of the celebrated Haldane and Semenov models for the sub-valleys Dirac cones, living on the domain walls. 
These findings open up a new direction towards the emergence and manipulation of the topological quantum valley, sub-valley Hall states, and possible application in valleytronics. Our theory can be generalized to many twisted bilayer systems, including twisted graphene, twisted $\alpha$-graphene, and twisted kagome bilayers at large-angle rotation.
\end{abstract}


\maketitle

{\it  Introduction.|} 
Small angle twisted bilayer graphene (TBG) has become an active research avenue after the discovery of various correlated states such as superconductor, Mott insulator, quantum fractal phase, etc~\cite{i1,i2,i3,i4,i5,i6,i7,i8,i9,i10,i11,i12,i13,i14,i15,i16,i17}. 
Many of these intriguing phenomena appear at certain rotation angles, coined \qq{magic angles} at which electronic kinetic energy is minimized \cite{i18,i19,i20,i21}. 
Interestingly, contrary to the early prediction of layer decoupling at large twist angles~\cite{i22}, recent studies have observed fascinating physical properties arising from interlayer coupling even at large rotation angles, including 
higher-order topology~\cite{i23,i24}, geometrical frustration~\cite{i25}, flat bands~\cite{i25}, hypermagic angles\cite{hypermagic}, and nontrivial optical response\cite{i26}.

The low-energy band structure of a large angle TBG is crucially affected by the higher-order Fourier components of the interlayer coupling~\cite{i27,i28}.
Explicitly, according to Ref.~\onlinecite{i27,i28}, commensurate angle TBGs can be classified into the sublattice exchange odd (SEO) and sublattice exchange even (SEE) configurations. 
SEE has only one stacking configuration similar to the AA stacked bilayer graphene (BLG), hence we call it the effective AA (EAA) structure.
The energy spectrum of EAA  possesses gapped band structures that host crystalline and higher-order topological insulators\cite{i23,i24}. 
On the contrary, SEO  can have two types of stacking structures similar to the Bernal-stacked BLG.
We name them the effective AB (EAB) and effective BA (EBA) configurations, respectively.
One fascinating phenomenon observed in AB/BA BLG is the quantum valley Hall effect (QVHE)\cite{di1,di2,di3,di4,di5,di6,di7,di8,di9,di10,di11,di12,di13,di16} . 
Considering the extensive research activities on QVHE and its great potential for device application,
it is highly desirable to extend related topological phenomena to other two-dimensional material setups. 
Especially, if such a  valley-dependent functionality can also be achieved in twisted BLG, 
the tunability of the twisted structure would facilitate further rapid development of valleytronics application.

In this work, we study the valley-dependent topological property of the TBG with a large twist angle.
In particular, we establish the \qq{bulk-edge}/\qq{bulk-domain} wall correspondence using theoretical models and realistic DFT-based $ab-initio$ computations (see supplemental material (SM) for the computational details).
We found that an out-of-plane electric field (E-field) opens up a topological gap in EAB/EBA TBG.
In the presence of $U(1)$ valley symmetry (i.e., without valley mixing), EAB/EBA TBG supports gapless domain wall (DW) modes between two distinct configurations with different valley Chern numbers (VCN).
The VCN (${C}_v$) is defined as integrating the Berry curvature around a given valley.
This QVHE and its DW modes can be tuned by applying an external E-field, twisting angle, and shifting one layer with respect to (w.r.t) another.
We found that the shifting works as an emergent E-field that competes with the external E-field to produce sub-valley Dirac cones and the corresponding sub-valley Hall effect. 
This emergent phenomenon is akin to the topological physics of the Haldane model in single-layer graphene. 
Namely, two sub-valleys experience a Haldane mass that emerges from the layer shifting and a Semenoff mass that is induced by an external E-field. 
The competition between these Haldane and Semenoff sub-valley masses gives rise to tunable DW modes in large-angle TBG.

\begin{figure}[t!]
	\centering
	\includegraphics[width=0.5\textwidth]{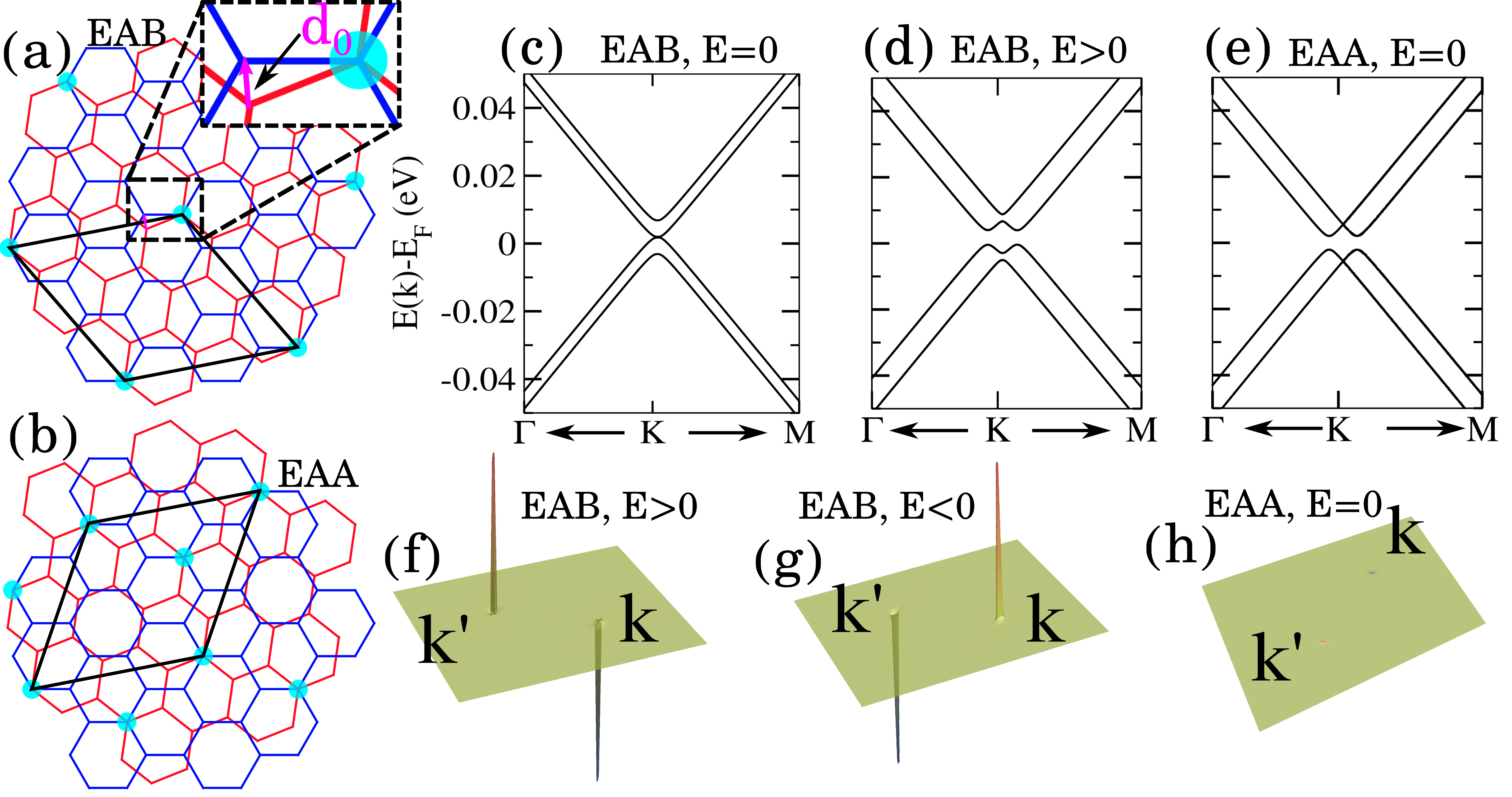}
	\caption{ The lattice structure of (a) EAB configuration at twist angle 38.21$^\circ$ and of (b) EAA configuration obtained by a shift ${\bf d} = \{0,2d_0\}$ from EAB.
 Blue and Red lines represent the top and bottom layers.
  Cyan discs represent the eclipsed atoms of the two layers. The primitive unit cell is shown by the solid black line.
	   Band structures of EAB structure (c) when $E = 0$ and (d) $E \neq 0$. $E\neq 0$ opens up a gap at K/K$^\prime$ valleys. 
	   (e) Band structure of EAA structure. Berry curvature distributions of EAB structure under (f) $E>0$ and (g) $E<0$. 
	   (h) Vanishing Berry curvature of EAA structure. }
	\label{fig1}
\end{figure}

{\it Structure and VCN}.| 
In the following, we consider a commensurate angle $\theta=38.21^\circ$. However, our conclusions also remain the same for other large commensurate angles.
The lattice structure of EAB and EAA configurations for $38.21^\circ$ is shown in Fig.~\ref{fig1}(a,b).
Note that EAA can be constructed from  EAB  by shifting one layer  w.r.t the other by a shift vector ${\bf d} = \{0,2d_0\}$.
The DFT band structure of EAB in the absence of an E-field (i.e, $E=0$) shows gapless quadratic bands touching at each valley K/K$^\prime$ of superlattice  Brillouin zone (see Fig.~\ref{fig1}(c)).
However, turning on the E-field gaps out those quadratic bands touching (see Fig.~\ref{fig1}(d)). 
 In Fig.~\ref{fig1}(f,g), we plot the Berry curvature distribution of EAB under $\pm E$ which shows that the gaped Dirac cones act as the hotspot of Berry curvature.
 The VCN changes its sign upon reversing the direction of the E-field or changing the valley, i.e. ${C}_v=\eta~sgn(E)$ (see SM for the derivation) where $\eta = \pm 1$ represents the valley index.
In contrast, the band structure of EAA configuration is always gapped irrespective of the presence or absence of $E$ (see Fig.~\ref{fig1}(e)).
 However, EAA does not show any Berry curvature because of the underlying $\mathcal{C}_{2z} \mathcal{T} $ symmetry, where $\mathcal{C}_{2z}: \{x,y,z\}\rightarrow \{-x,-y,z\}$ and $\mathcal{T}$ are  two-fold rotation and time reversal symmetries, respectively. This is because $\mathcal{C}_{2z} \mathcal{T}$ symmetry enforces the reality of the wave functions and vanishing Berry curvature\cite{bj1,bj2,bj3}.

\begin{figure}[t!]
	\centering
	\includegraphics[width=0.5\textwidth]{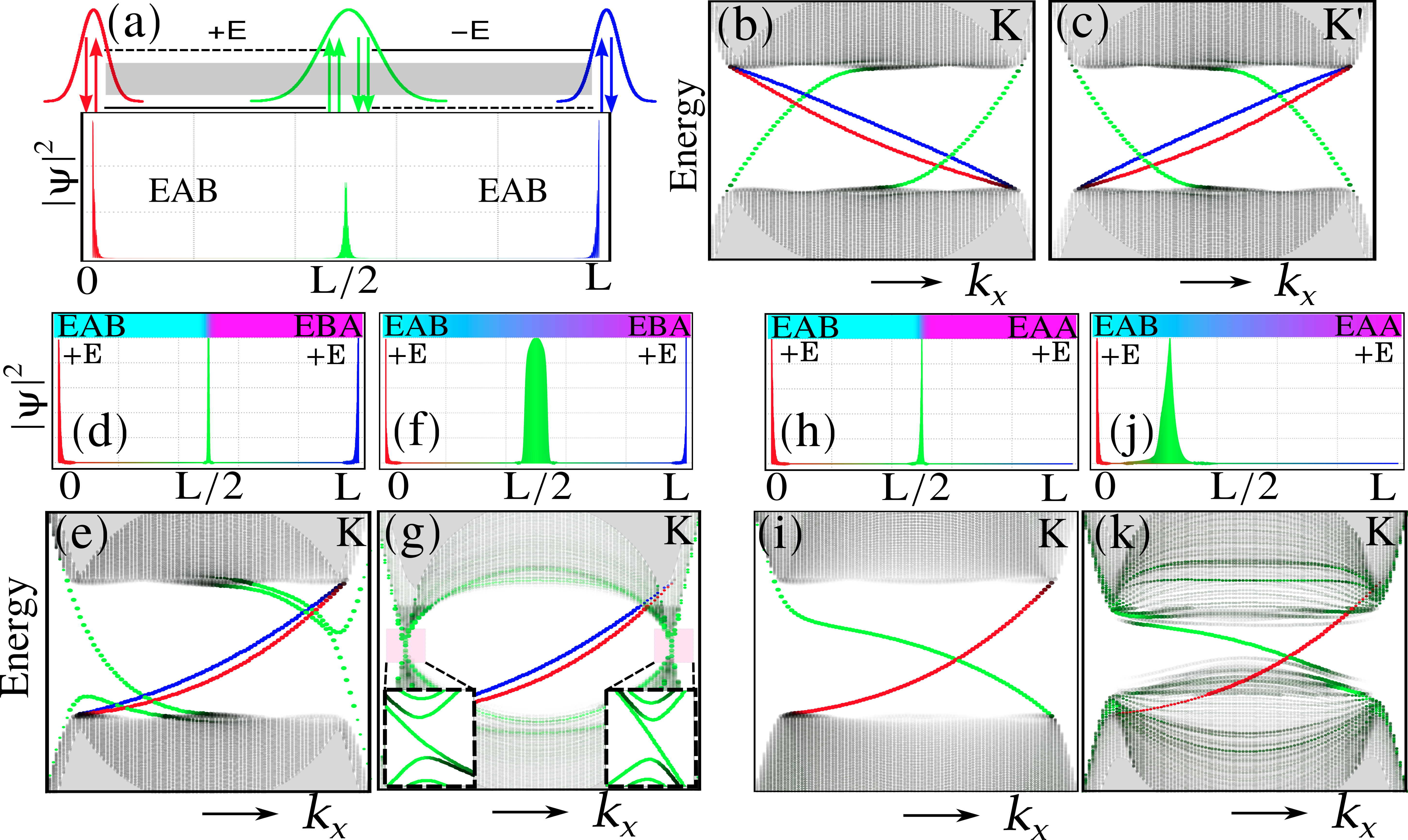}
	\caption{Edge and DW modes under different E-field and DW configurations.
	 The green (blue/red) color represents the DW (edge) modes. 
	 (a) A DW is created in EAB by changing the sign of $E$ (i.e. $+E$ and $-E$) across the system.
	  Wave functions are strongly localized at the DW ($L/2$) and edges ($0$ and $L$).
	  (b,c) Energy dispersion along DW and edges near each valley  K/K$^\prime$ for a ribbon geometry with periodicity along the DW.
      DWs are created (d) sharply and (f) smoothly between the EAB and EBA (i.e. alternate stacking configuration) under $+E$. Corresponding localized modes are shown in (e,g) for K valley.
      (h) Sharp and (j) smooth DW are created between the EAB and EAA under $+E$.
        For smooth junction in (j), the wave function is localized near EAB. 
        Corresponding edge and DW states are shown in (i,k) for K valley.
	    The color bar in (d),(f),(h), and (j) represent how the DW are created, sharply or smoothly, between two different configurations as represented by cyan and magenta colors.}
	\label{fig2}
\end{figure}

{\it Domain Wall modes.|}
In general, gapless modes appear at the edge or DW between two systems with distinct topological invariants. 
As such, we expect the edge/DW modes to appear at the edges or DW between two configurations with different VCNs.
Also, opposite VCN from opposite valleys enforces the edge/DW modes to propagate with opposite velocities. 
Furthermore, the absolute change of VCN, $\Delta \mathcal{C}_v=|\mathcal{C}^\text{A}_v-\mathcal{C}^\text{B}_v|$ between two regions A and B is equal to the number of DW modes between those regions.
For instance, in Fig.~\ref{fig2}(a), we consider a TBG ribbon with EAB configuration. We then apply the opposite E-field on the two sides of the system. 
In this case, as $\Delta \mathcal{C}_v=\pm 2$ for the K and K$^\prime$ valleys, respectively, the system hosts two counter-propagating modes in each valley. 
We numerically confirm the existence of DW and edge modes in the energy spectrum calculation where we assume the periodicity along the boundary between two regions, as shown in Fig.~\ref{fig2}(b,c) by the green and red/blue colors (See SM for details of tight-binding calculations).
We plot the corresponding wave functions in Fig.~\ref{fig2}(a), which are exponentially localized at the DW and edges, respectively.
The existence of edge states can be explained by the change of VCN between EAB/EBA and the vacuum. 
For instance, EAB obtains $|\mathcal{C}_v|=1$ under the E-field, which gives $\Delta \mathcal{C}_v=1 $ between EAB and the vacuum. 
Note that the change of VCN between EAB and vacuum is of the same sign for both the left and right edges. 
This is because the sign of VCN changes on two sides by the opposite sign of the E-field. 
Therefore, the states at the two edges obtain the same propagation direction.
Furthermore, as the change of VCN between EAB and EBA  under the same E-field is equal to $\Delta \mathcal{C}_v=\pm 2$, we expect similar edge and DW modes localized between them (see Fig.~\ref{fig2}(d,e)).
The existence of DW modes is very robust as their appearance does not depend on the detailed chemistry of the DW. 
For demonstration, we construct a ribbon of TBG and then slowly shift one layer w.r.t the other layer, to construct a smooth DW between  EAB and EBA (see Fig.~\ref{fig2} (f)).
We observe two in-gap DW modes and edge modes as shown in the inset of Fig.~\ref{fig2} (g), similar to Fig.~\ref{fig2} (e).

\begin{figure}[t!]
	\centering
	\includegraphics[width=0.5\textwidth]{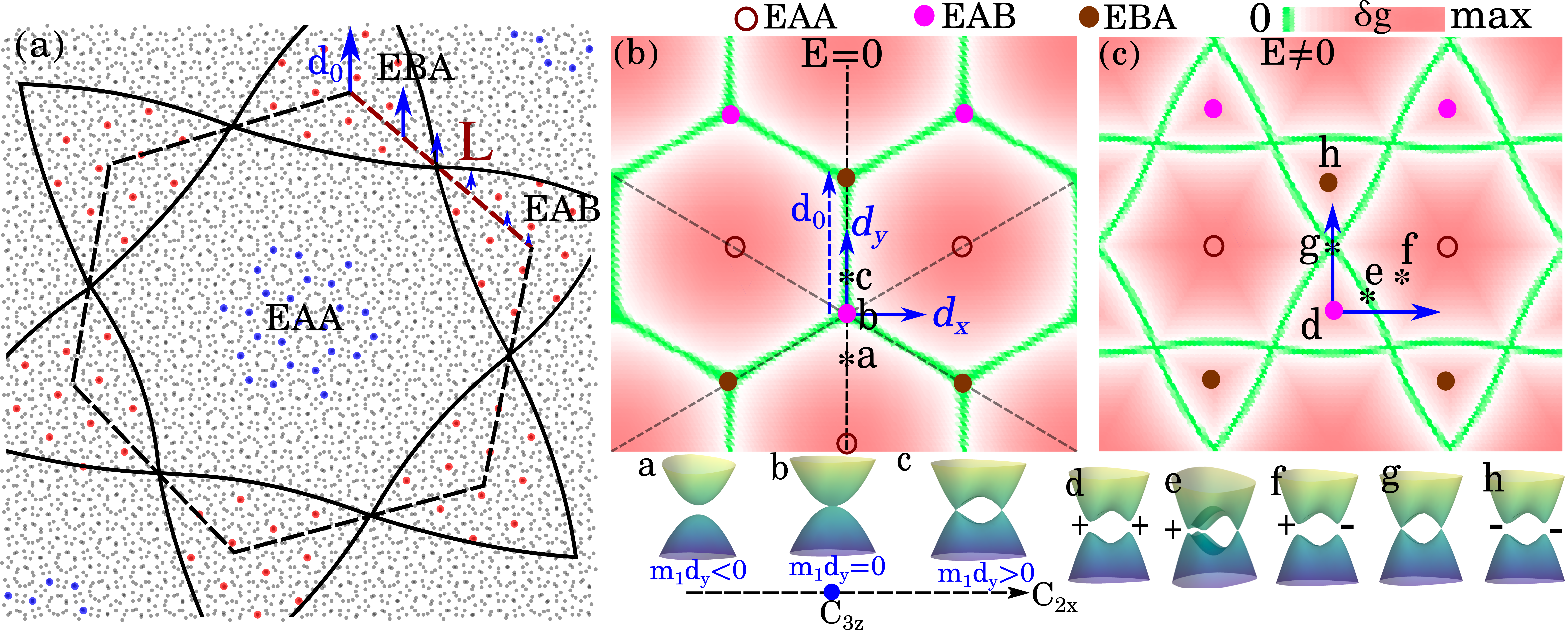}
	\caption{ (a) TBG with a small twisting ($\delta \theta_c$) from the commensurate angle $\theta_c = 38.21^\circ$.
 The local configurations of EAA and EAB/EBA (separated by a black solid line) can be traced by tracking the coordination of eclipsed atoms (atoms in two layers stacked on top, shown by red and blue dots) which form a triangular
(hexagonal) coordination at the EAA (EAB/EBA) region.
 Map of the direct energy gap ($\delta \mathbf{g}$) of local structures in real space as a function of the shift vector {\bf d} for (b) $E=0$, and (c) $E\neq 0$.  
 The vector {\bf d$_0$} in (b) has a unique correspondence with $L$ in (a) as shown by the increasing {\bf d$_0$} (blue arrows) along the line with the length $L= l/[2~\text{sin}(\delta \theta_c/2)]$ in (a) where \textit{l} represents the lattice constant of EAA/EAB.
 The local structures with different {\bf d} vectors are labeled by a,b,c in (b) and d,e,f,g,h in (c).
 The energy dispersions in momentum space near K/K$^\prime$ corresponding to the local structures are shown in the bottom panel of (b) and (c).  
 The $\pm$ symbols represent the sign of the Berry curvature of gapped Dirac cones.}  
	\label{fig3}
\end{figure}

The existence of DW modes in EAB/EBA configuration is similar to the case of AB stacked BLG.
However, in contrast to gapless AA stacked  BLG,  EAA configuration of commensurate TBG is gapped, and $C_{v}=0$.
Therefore, one can imagine EAA effectively behaves as a vacuum state.
Consequently, by creating a DW between EAA and EAB, we expect the appearance of DW modes between them. 
In Fig.~\ref{fig2}(h,i), indeed, we confirm the existence of one DW mode at the center of the system. 
Additionally, as it is shown in Fig.~\ref{fig2}(h,i), only the edge located on the EAB side support edge modes, while the EAA edge does not support any edge modes. 
This is in accordance with the nonzero (zero) VCN of EAB (EAA). 
Similar to the EAB-EBA setup in Fig.~\ref{fig2}(f), a DW between EAB and EAA, created by smoothly changing the local configuration, also hosts DW modes as shown in Fig.~\ref{fig2}(j,k). 
However, the position of the DW mode wave functions is located closer to the EAB region (see Fig.~\ref{fig2}(j)).

{\it Incommensurate structure and domain wall network.|}
When two layers are slightly rotated from a commensurate angle $\theta_c$, it generally breaks the commensurability of the lattice over a large length scale in the lattice (see SM). 
However, such an incommensurate system locally resembles a commensurate configuration (see Fig.~\ref{fig3}(a)) that can be constructed by a certain amount of shifting of one layer w.r.t the other layer, where the nearby local structures are connected by smooth shifting \cite{hypermagic}. 
Moreover, the local structural environment could be large enough to construct momentum space locally\cite{i18}.
In the following, we study the existence of DW modes based on the energy spectrum of the local structures. 
First, we construct the commensurate TBG Hamiltonian at $\theta_c = 38.21^\circ$, and then the  Hamiltonian of the local structure is constructed by using the appropriate shift vector $\bf d$. 
In TBG, the effective low-energy Hamiltonian comprises two parts.
The intralayer part can be represented by the Dirac Hamiltonian $H_{1,2} = -i\nu (\sigma_x k_x+\sigma_y k_y) $ where $\nu$ is the Fermi velocity, and $\sigma_i$ represents the Pauli matrices.
The interlayer coupling part Hamiltonian for $\theta_c = 38.21^\circ$ is written as, (see SM for the derivation)
 \begin{equation} \label{eq2}
 H_{\bf k,\tilde k}^{\alpha \beta} = t_{\bf k+G} \sum_{i=1}^3 e^{i \bf G_i \cdot (\tau^\alpha + d)} e^{-i \bf  \tilde G_i \cdot\tau^\beta} 
 \end{equation}
 where $\alpha, \beta$ represent the sublattice index with positions $\tau_i (i=\alpha, \beta)$, ${\bf G} = \sum {n_i {\bf b_i} } $ and ${\bf \tilde{G} } = \sum{ \tilde{n_i} {\bf \tilde{b_i} } } $ are the reciprocal lattice vectors for the two layers, respectively, $t_{\bf k+G}$ is the Fourier coefficient of the tight-binding hopping in the momentum space whose amplitude decay rapidly with $|{\bf k+G}|$. 
 The ${\bf G}$ and ${\bf \tilde{G}} $ vectors that satisfy the momentum conservation process with the largest $|t|$ is given as,
($\bf G_1$, $\bf \tilde{G}_1) = (b_1, \tilde{b}_1 + \tilde{b}_2 )$, 
($\bf G_2$, $\bf \tilde{G}_2) = (- 2b_1 - 2b_2, -\tilde{b}_1 - 2\tilde{b}_2 )$, and
($\bf G_3$, $\bf \tilde{G}_3) = (- b_1 + b_2, -2\tilde{b}_1 )$. 
The Hamiltonian for the external E-field is written as $H_E = \text{diag}(E,E,-E,-E)$.

In Fig.~\ref{fig3}(b,c), we show the map of the direct energy gap ($\Delta g$) of the local structures for $E=0$ and $E\neq 0$ as a function of the shift vector {\bf d}.
When $E=0$, some local structures possess two important symmetries. 
They are two-fold in-plane ($\mathcal{C}_{2}$)  and three-fold out-of-plane  ($\mathcal{C}_{3z}$) rotational symmetries. 
For instance, let us consider a locus of local structures on a line, connecting EAA, EAB, and EBA (vertical dashed line in Fig.~\ref{fig3}(b)). 
For $E=0$, all the local structures on this line are $\mathcal{C}_{2}$ symmetric with an additional $\mathcal{C}_{3z}$ at the positions EAB and EBA. 
 Figure~\ref{fig3}(b) shows that the local structure at EAA is gaped which is also consistent with our DFT calculations, as discussed in Fig.~\ref{fig1}(e).
However, from the EAA towards the EAB (on the dashed line), the $\Delta g$ decreases and finally closes at EAB at which quadratic band crossing appears at K/K$^\prime$ valley.
The energy dispersion near a valley, corresponding to the local structure in positions \qq{a} and \qq{b} is shown in the bottom panel of Fig.~\ref{fig3}(b).
By further moving towards EBA from EAB, the quadratic band touching split into two Dirac cones, which we call sub-valley Dirac cones (see the energy dispersion at a general position \qq{c} in the bottom panel of Fig.~\ref{fig3}(b)).  
These sub-valley Dirac cones are protected by the underlying $\mathcal{C}_{2}$ symmetry and finally merge together at EBA to form a quadratic band touching.

For a better understanding of the DW physics, we derive a low-energy Hamiltonian up to the second order in the momentum under an external E-field considering a slight shift from the EAB on the vertical dashed line (i.e. d$_x$=0), which becomes (see SM for derivation)
\begin{equation} \label{eq4}
\begin{aligned}
H_{0,d_y}(k_x,k_y)= &
(\alpha (k_x^2 - k_y^2) + m_1 d_y) )\sigma_x - 2\alpha ~ k_x k_y \sigma_y \\
& + (E+m_3 ~ d_y k_y) \sigma_z, 
\end{aligned}
\end{equation}
where $\alpha, m_i (i=0,1,2,3)$ are the constants.
When $E=0$, the system recovers $\mathcal{C}_{2}$ = $\sigma_x$ symmetry and it support gapless spectrum at $k_y=0$ and $k_x = \pm k_0$, where $k_0=\sqrt{m_1 d_y}$.
As such, for $m_1>0$, two gapless solutions appear only when $d_y>0$ and no solutions for $d_y<0$. 
This is in accordance with Fig.~\ref{fig3}(b), where gapless configurations only appear for $d_y>0$, while the local structures for $d_y<0$ are gaped. The two regions meet each other at EAB ($d_y=0$) on the vertical dashed line.
The quadratic band touching at this point is enforced by the underlying $C_{3z} = e^{i \frac{2\pi}{3} \sigma_z }$ symmetry. 

Now, we write an effective Hamiltonian for each sub-valley Dirac cone (for $d_y>0$) as $H_{\text{sv}} = k_x \sigma_x + k_y \sigma_y$ sitting at $\pm k_0$. 
These gapless Dirac cones cannot be gapped out unless the $\mathcal{C}_{2}$ symmetry is broken by either a nonzero shift along $d_x$ or an E-field $E$. 
The $d_x$ and $E$ result in a mass term for these sub-valley Dirac cones,
\begin{equation}\label{eq5}
    \mathcal{M}_{\text{sv}}=(E \pm d_x k_0)\sigma_z,
\end{equation}
where the first mass term comes from $E$ and the second term is related to $d_x$ which can be interpreted as an emergent electric field.
The two sub-valley Dirac cones get the same sign of masses by $E$, akin to the Semenoff mass as in the case of single-layer graphene \cite{semenoff}. 
However, away from the $\mathcal{C}_{2}$ symmetric DW configurations (i.e. $d_x \neq 0$), the second term in Eq.~(\ref{eq5}) induces the opposite sign of masses to the two sub-valley Dirac cones, resembling the Haldane mass as in single layer graphene \cite{haldane}.
This sub-valley Haldane mass gives opposite sub-valley Chern numbers [sub-VCN] for two gaped sub-valley Dirac cones. 
The sub-VCN is defined by the integration of the Berry curvature around each gaped sub-valley Dirac cone.
Note that the sign of the sub-VCN changes when $d_x$ changes its sign about $d_x=0$ line. 
For instance, if sub-VCN for two gapped Dirac cones are (+1/2,-1/2) for $d_x >0$, it is then (-1/2,+1/2) for $d_x < 0$.
Therefore, the change of sub-VCN between $d_x <0$ and $d_x > 0$ is $\pm 1$ that protects the two counter-propagating DW modes at $d_x =0$. 
Therefore this system can host a quantum sub-valley Hall effect even in the absence of a true external E-field.

In the presence of $E$, the $\mathcal{C}_{2}$ symmetry is broken which gaps out both the sub-valley Dirac cones as depicted in position \qq{d} and \qq{h} in Fig.~\ref{fig3}(c).
According to Eq.~(\ref{eq5}), as $d_x$ increases, one Dirac cone remains gapped while the other Dirac cone goes through the gap closing and reopening at the DW when $|E| = |d_x k_0|$ (see \qq{e} and \qq{f} in Fig.~\ref{fig3}(c)). 
This gapless Dirac point corresponds to a DW mode which is also consistent with $\Delta C_v = 1$ between the positions \qq{d} and \qq{f} in Fig.~\ref{fig3}(c).
The phase transition resembles the Haldane model with a different sub-lattice potential (known as Semenoff mass) in the case of single-layer graphene where the competition between Semenoff and Haldane mass dictated the topological phase boundary.
Furthermore, at the EAA, two gapped sub-valley Dirac cones meet each other and annihilate their underlying Berry curvature and hence the sub-VCN.
Note that, the $\Delta C_v = 2$ between the positions \qq{d} and \qq{h} (EAB and EBA) in Fig.~\ref{fig3}(c) which ensures two DW modes, represented by two gapless Dirac cones at position \qq {g} (see bottom panel of Fig.~\ref{fig3}(b)). 
The DW mode between  EAB/EBA and EAA in Fig.~\ref{fig3}(c) appears closer to the EAB/EBA region which is consistent with our tight-binding calculation in Fig.\ref{fig2}(j).
On the contrary, the gapless DW configuration between EAB and EBA (Fig.~\ref{fig3}(b)) appears exactly between them which is consistent with Fig.~\ref{fig2}(f).

\begin{table}[t!]
\caption{\label{table1} Number of atoms (N), in-plane lattice constant (\textit{l}), and band gap ($\Delta$) of the EAA configuration for several $\theta_c$ with largest layer coupling at large-angle of rotation.   }
\begin{ruledtabular}
\begin{tabular}{ccccc}
$\theta_c$ &  N & \textit{l} in \AA & $\Delta$ in meV  \\
\hline
$38.21^\circ$  &    28   &  6.46   &   5    \\
$32.2^\circ$   &    52   &  8.85   &   0.1   \\
$43.17^\circ$  &    76   &  10.7   &   0.2    \\
$42.1^\circ$   &    124  &  13.67  &   0.25    \\
\end{tabular}
\end{ruledtabular}
\end{table}

{\it Discussions.|}
Let us compare QVHE in large-angle TBG to that in untwisted or small-angle twisted TBG.
First, in the presence of an E-field, all the configurations EAA, EAB, and EBA are gapped in large-angle commensurate TBG which makes it distinct from the small-angle twisting where AA configuration is gapless even under E-field.   
Second, unlike the small-angle case, the relaxation effect is insignificant in large-angle twisting. 
It is because the ground state energy difference (computed using DFT) between AA and AB/BA is $\sim$103 meV/atom whereas it is only 0.14 meV/atom in the case of EAA and EAB/EBA at $\theta_c = 38.21^\circ$. 
Therefore lattice relaxation makes the AA region shrink into a topological point defect \cite{di16} in the small-angle TBG. 
However, in large-angle TBG, EAA, and EAB/EBA all exist as stable local structures and support proposed DW states.

Although our theory is generally valid, it is experimentally more significant near few $\theta_c$ as the band gap ($\Delta$) decreases rapidly when the lattice constant (\textit{l}) and the number of atoms (N) of the commensurate unit cell increase as listed in Table~\ref{table1}
where we show $\Delta$ for four $\theta_c$ with smallest \textit{l} and N.

 The proposed DW modes can be detected both in STM and transport measurements.
As the size of DW network is larger than STM resolution, it can be detected by the STM local density of states and dc optical conductance measurement~\cite{di10}.
Moreover, the 2D conducting DW network pattern can induce intriguing non-Fermi liquid behaviour above a critical temperature $T_x \sim \frac{\hbar v_f}{k_B L}$ where $v_f$, $\hbar$, $k_B$, $L$ represent the DW Fermi velocity, reduced Planck constant, Boltzmann constant, length of 1D conducting channels, respectively~\cite{transport}.
In our case, for a rotation $\delta \theta_c =0.05^\circ$ from $\theta_c= 38.21^\circ$, $L$ is $\sim$740 nm, and computed $v_f$ is $\sim$0.6$\times$10$^6$ m/s which gives $T_x\approx 7K$. 
As the band gap of EAA is $\sim$5 meV ($\sim$58K in temperature scale), the 2D DW network at $\delta \theta_c =0.05^\circ$ is expected to show 1D non-Fermi liquid transport from 7K up to $\sim$58K.

Finally, we note that our theoretical idea can also be applied to other related structures such as twisted bilayer $\alpha$-graphene~\cite{di14} and twisted kagome bilayers~\cite{di15} (see SM for more detail) with more tunability.

\begin{acknowledgements}
We thank Mikito Koshino and Joonho Jang for their stimulating discussions. 
C.M. thanks Dr. Soumya Datta for the insightful discussion on STM spectroscopy.
C.M., R.G., and B.J.Y. were supported by the Institute for Basic Science in Korea (Grant No. IBS-R009-D1),
Samsung  Science and Technology Foundation under Project Number SSTF-BA2002-06,
the National Research Foundation of Korea (NRF) grant funded by the Korean government (MSIT) (No.2021R1A2C4002773, and No. NRF-2021R1A5A1032996).
\end{acknowledgements}




\end{document}